\documentclass[aip,apl]{revtex4-1}

\usepackage{graphicx}
\usepackage{dcolumn}
\usepackage{bm}

\usepackage[utf8]{inputenc}
\usepackage[latin9]{luainputenc}
\usepackage[T1]{fontenc}
\usepackage{mathptmx}
\usepackage{etoolbox}
\usepackage{amsfonts}
\usepackage{amsmath}
\usepackage{textcomp}
\usepackage{amstext}
\usepackage{subscript}
\usepackage{graphics}
\usepackage{float}
\usepackage{color}
\usepackage{pifont}
\usepackage{pmboxdraw}
\usepackage{amssymb}
\usepackage{esint}
\usepackage{tabularx}
\usepackage[export]{adjustbox}
\usepackage{hyperref}
\usepackage{upgreek,textgreek}
\usepackage{siunitx}
\usepackage{multirow}
\usepackage{mhchem}
\usepackage[section]{placeins}

\makeatletter
\def\@email#1#2{%
 \endgroup
 \patchcmd{\titleblock@produce}
  {\frontmatter@RRAPformat}
  {\frontmatter@RRAPformat{\produce@RRAP{*#1\href{mailto:#2}{#2}}}\frontmatter@RRAPformat}
  {}{}
}%
\makeatother
\begin{document}

\preprint{AIP/123-QED}

\title{Large Nernst Effect in a layered metallic antiferromagnet EuAl$_2$Si$_2$}

\author{Kunya Yang}
\email{kunyayang@cqu.edu.cn}
\affiliation{Low Temperature Physics Lab, College of Physics \& Center of Quantum Materials and Devices, Chongqing University, Chongqing 401331, China}
\affiliation{Department of Physics, Chongqing Three Gorges University, Chongqing 404100, China}

\author{Wei Xia}
\affiliation{School of Physical Science and Technology, ShanghaiTech University, Shanghai 201210, China}
\affiliation{ShanghaiTech Laboratory for Topological Physics, Shanghai 201210, China}

\author{Xinrun Mi}
\affiliation{Low Temperature Physics Lab, College of Physics \& Center of Quantum Materials and Devices, Chongqing University, Chongqing 401331, China}
\author{Yiyue zhang}
\affiliation{Low Temperature Physics Lab, College of Physics \& Center of Quantum Materials and Devices, Chongqing University, Chongqing 401331, China}
\author{Long zhang}
\affiliation{Low Temperature Physics Lab, College of Physics \& Center of Quantum Materials and Devices, Chongqing University, Chongqing 401331, China}

\author{Aifeng Wang}
\affiliation{Low Temperature Physics Lab, College of Physics \& Center of Quantum Materials and Devices, Chongqing University, Chongqing 401331, China}
\author{Yisheng Chai}
\affiliation{Low Temperature Physics Lab, College of Physics \& Center of Quantum Materials and Devices, Chongqing University, Chongqing 401331, China}

\author{Xiaoyuan Zhou}
\affiliation{Low Temperature Physics Lab, College of Physics \& Center of Quantum Materials and Devices, Chongqing University, Chongqing 401331, China}

\author{Yanfeng Guo}
\email{guoyf@shanghaitech.edu.cn}
\affiliation{School of Physical Science and Technology, ShanghaiTech University, Shanghai 201210, China}
\affiliation{ShanghaiTech Laboratory for Topological Physics, Shanghai 201210, China}

\author{Mingquan He}
\email{mingquan.he@cqu.edu.cn}
\affiliation{Low Temperature Physics Lab, College of Physics \& Center of Quantum Materials and Devices, Chongqing University, Chongqing 401331, China}

\date{\today}

\begin{abstract}
 The large Nernst effect is advantageous for developing transverse Nernst thermoelectric generators or Ettingshausen coolers within a single component, avoiding the complexity of electron- and hole-modules in longitudinal Seebeck thermoelectric devices. We report a large Nernst signal reaching 130 $\upmu$V/K at 8 K and 13 T in the layered metallic antiferromagnet EuAl$_2$Si$_2$. Notably, this large transverse Nernst thermopower is two orders of magnitude greater than its longitudinal counterpart. The Nernst coefficient peaks around 4 K and 8 K at 3 T and 13 T, respectively. At similar temperatures, both the Hall coefficient and the Seebeck signal change sign. Additionally, nearly compensated electron- and hole-like carriers with high mobility ($\sim$ 4000 cm$^2$/Vs at 4 K) are revealed from the magnetoconductivity. These findings suggest that the large Nernst effect and vanishing Seebeck thermopower in EuAl$_2$Si$_2$ are due to the compensated electron- and hole-like bands, along with the high mobility of the Weyl band near the Fermi level. Our results underscore the importance of band compensation and topological fermiology in achieving large Nernst thermopower and exploring potential Nernst thermoelectric applications at low temperatures.

\end{abstract}

\maketitle
The thermoelectric effect, which characterizes the flow of entropy in charge carriers within a conductor, enables the conversion between thermal and electrical energy. Efficient thermoelectric power generators and coolers are crucial for waste heat recovery and solid-state refrigeration \cite{Goldsmid, thermoele2013, thermoele2017}. Conventional thermoelectric devices rely on the longitudinal thermoelectric effect, specifically the Seebeck effect, and typically involve both electron- and hole-type materials\cite{comprehensive2020, thermoelectric2018}. This can lead to contact resistance and compatibility issues when connecting different materials, thereby limiting device performance\cite{contact2020,contact2024}. In contrast, the transverse thermoelectric effect, or Nernst effect, allows for additive contributions from electron- and hole-like carriers within a single material, making it a promising avenue for thermoelectric applications\cite{Pan2022,xiang2020large,Watzman_NbP,ZrTe5}. However, the Nernst signal is usually several orders of magnitude smaller than the Seebeck effect, constraining its use. Therefore, exploring materials with a large Nernst effect is of significant fundamental and technological importance.

In the presence of a magnetic field ($\mathbf{B}\parallel\mathbf{z}$), the Nernst signal ($ S_{yx}=\frac{E_y}{\nabla_x T}$) is manifested by a transverse electric field ($\mathbf{E}\parallel\mathbf{y}$) produced by a longitudinal temperature gradient ($-\nabla T\parallel\mathbf{{x}}$) within a solid. The Nernst effect was first discovered by Ettingshausen and Nernst in bismuth \cite{Nernst}, which still holds the record for the highest Nernst coefficient ($\nu=S_{yx}/B\sim 6$ mV/ KT) in metals  \cite{Behnia_Bi}. The exceptionally high Nernst signal in bismuth originates from its high mobility ($\mu\sim10^7$ cm$^2$/Vs) and low Fermi energy ($E_\mathrm{F}\sim10$ meV) as the Nernst coefficient in a semi-classical picture is described by  \cite{behnia2009,behnia2016}:
\begin{equation}
    \nu\sim\frac{\pi^2}{3}\frac{k^2_\mathrm{B}T}{e}\frac{\mu}{E_\mathrm{F}}=283\frac{\mu}{E_\mathrm{F}}T,
    \label{eq1}
\end{equation}
where $e$ and $k_\mathrm{B}$ are the electric charge and Boltzmann constant. This relation applies well to various metals, ranging from elemental to strongly correlated materials \cite{behnia2009,behnia2016} [see Fig. \ref{fig:3}(e)], thus offering a prominent guide for searching materials with a large Nernst effect. In this context, topological semimetals with linearly dispersing bands are excellent candidates for achieving high mobility and large Nernst thermopower \cite{Zhu_WTe2,Pan2022,XuYbMnSb,Watzman_NbP}. Furthermore, electron-hole compensation, often found in linearly dispersive bands, can further enhance the Nernst signal by avoiding the Sondheimer cancellation, which leads to a vanishing Nernst signal commonly observed in simple metals \cite{Sondheimer,Wangyayu2001,Pan2022,XuYbMnSb,Wu_NbAs2}. Indeed, large Nernst effects have been reported in various topological semimetals, as shown in TABLE \ref{Tab1}. Still, non-ideal electron-hole compensation and the presence of trivial bands often lead to large Seebeck signals. Observing a transverse Nernst effect that is much larger than the longitudinal Seebeck effect is rare.

\begin{table}
\centering
\caption{Large Nernst Thermopower in Various Topological Semimetals. This table summarizes the exceptionally large Nernst thermopower reported in several topological semimetals. The values include the peak Nernst signal, the temperature at which it occurs, and the applied magnetic field.
\label{parameter_of_specific_heat}}
\renewcommand*{\arraystretch}{1}
\begin{tabularx}{1\textwidth}{
>{\centering\arraybackslash}X
>{\centering\arraybackslash}X 
>{\centering\arraybackslash}X
>{\centering\arraybackslash}X}
\hline
\hline
Materials & $S_{yx}$ ($\upmu$V/K) & $T$ (K)  &$B$ (T)\\
\hline
\ce{WTe2}\cite{Zhu_WTe2,Pan2022} & 7000 & 11 & 9\\
\ce{TaP}\cite{TaP2020} & 1070 & 40 & 9\\
NbP\cite{Watzman_NbP} & 800  & 109 & 9\\
\ce{NbSb2} \cite{Li2022_NbSb2} & 600 & 25 & 9\\
\ce{NbAs2}\cite{Wu_NbAs2} & 600 & 35 & 9\\
\ce{KMgBi}\cite{Ochs_KMgBi} & 200 & 80 & 1\\
\ce{Mg2Pb}\cite{Mg2Pb2021} & 190 & 30 & 9\\
\ce{KMn6Bi5}\cite{DongKMn6Bi5} & 150 & 5 & 9\\
\ce{Cd3As2}\cite{xiang2020large} & 85 & 350 & 9 \\
\ce{YbMnSb2}\cite{XuYbMnSb} & 40 & 120 & 14 \\
\ce{PtSn4}\cite{PtSn2020} & 39 & 15 & 9\\
\hline
\hline
\end{tabularx}
\label{Tab1}
\end{table}

In this article, we report a large Nernst signal with a nearly vanishing Seebeck effect in the layered antiferromagnetic metal \ce{EuAl2Si2}. The Nernst thermopower peaks around 8 K in a 13 T magnetic field, reaching a substantial value of  $S_{yx}\sim130$ $\upmu$V/K.  In contrast, the Seebeck signal is exceedingly small  ($S_{xx}\sim1$ $\upmu$V/K, 8 K, 13 T) and changes sign at approximately 10 K. These unusual properties arise from electron-hole compensation, characterized by high mobilities ($\sim$ 4000 cm$^2$/Vs) and a small Fermi energy  ($\sim$ 9 meV).   

High-quality single crystalline \ce{EuAl2Si2} samples were synthesized using the self-flux method \cite{xia2023}. Single crystal X-ray diffraction data were collected with a Bruker D8 Venture diffractometer. Seebeck and Nernst measurements were conducted using a steady-state method in a 14 T Oxford cryostat. The Seebeck and Nernst voltages were recorded with Keithley 2182A nanovoltmeters. Magnetization, specific heat, and magnetoconductivity measurements were performed using a Physical Property Measurement System (Quantum Design Dynacool 9 T).

\begin{figure*}
\centering
\includegraphics[scale=0.22]{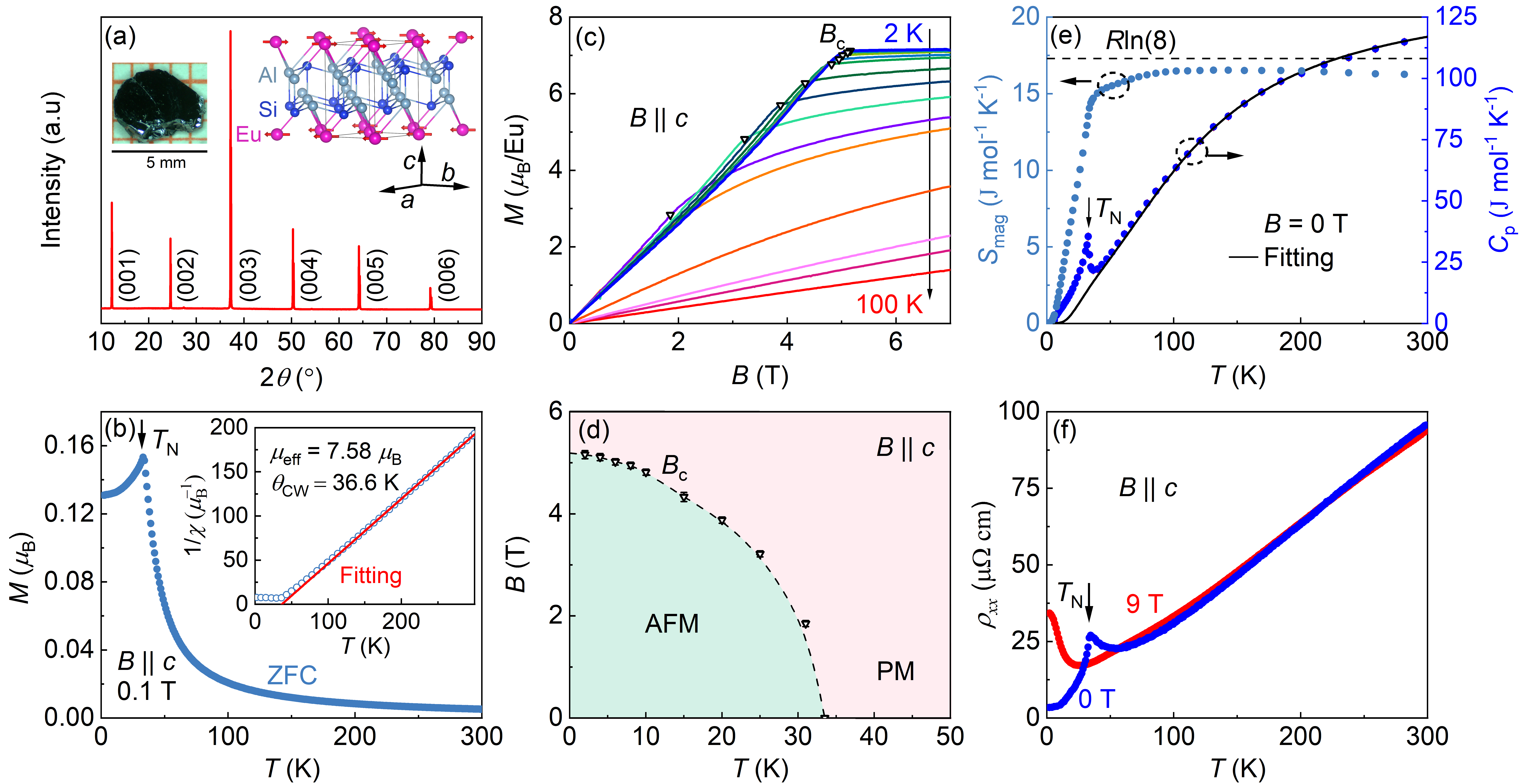}
\caption{ (a) X-ray diffraction pattern of a \ce{EuAl2Si2} single crystal. The upper right inset shows the crystal structure of EuAl$_2$Si$_2$ and the upper left inset presents an optical photo of a typical  \ce{EuAl2Si2} single crystal. (b) and (c) Temperature-dependent ($M-T$) and isothermal magnetization ($M-B$) with $B\parallel c$. The AFM transition is identified at $T_\mathrm{N}=33.5$ K.  The inset in (b) shows a Curie-Weiss fitting (red solid line) of the susceptibility. (d) $B-T$ phase diagram derived from the data in (c). (e) Zero-field specific heat ($C_\mathrm{p}$) and magnetic entropy ($S_\mathrm{mag}$) of \ce{EuAl2Si2}. The black solid line represents Debye-Einstein fitting of the phonon background. (f) Temperature dependence of longitudinal resistivity ($\rho_{xx}$).  }
\label{fig1}
\end{figure*}

\ce{EuAl2Si2} adopts the \ce{CaAl2Si2}-type structure (trigonal, $P\overline{3}m1$, No. 164), as illustrated in the inset of Fig. \ref{fig1}(a). This structure features two layers of Al-Si zigzag chains sandwiched between two layers of Eu. Below $T_\mathrm{N} \approx 33.5$ K, \ce{EuAl2Si2} transitions into an A-type antiferromagnetic (AFM) state, characterized by Eu moments aligned ferromagnetically within the $ab$-plane and antiferromagnetically along the $c$-axis \cite{schobinger1989, kranenberg2000, maurya2015, parfenov2021high, xia2023,Chen_EuAlSi}. The main panel of Fig. \ref{fig1}(a) displays the X-ray diffraction pattern of a typical \ce{EuAl2Si2} single crystal. The well-resolved ($00L$) peaks confirm the pure phase and [001] preferred orientation of the as-grown crystal. The AFM transition is evident at $T_\mathrm{N} = 33.5$ K, as indicated by a peak in the temperature-dependent magnetization shown in Fig. \ref{fig1}(b) with a magnetic field applied along the $c$-axis ($B \parallel c$). A Curie-Weiss fit of the susceptibility above 100 K yields an effective moment of $\mu_\mathrm{eff} = 7.58$ $\mu_\mathrm{B}$ and a Curie-Weiss temperature of $\theta_\mathrm{CW} = 36.6$ K [see Fig. \ref{fig1}(b) inset], consistent with previous reports \cite{schobinger1989, kranenberg2000, maurya2015, parfenov2021high, xia2023,Chen_EuAlSi}. This effective moment is close to the theoretical value of 7.94 $\mu_\mathrm{B}$ for Eu$^{2+}$. The positive Curie-Weiss temperature indicates predominant ferromagnetic (FM) fluctuations due to intralayer FM interactions in the paramagnetic state. For $B \parallel c$, a field-polarized FM state with a saturated moment of approximately 7.15 $\mu_\mathrm{B}$/Eu is achieved above a critical field $B_\mathrm{c} \approx 5.16$ T at 2 K [Fig. \ref{fig1}(c)]. Two critical fields with lower values have been observed for $B \parallel ab$ \cite{schobinger1989, kranenberg2000, maurya2015, parfenov2021high, xia2023,Chen_EuAlSi}, indicating an easy-plane anisotropy consistent with the magnetic structure shown in Fig. \ref{fig1}(a). Using the isothermal magnetization data shown in Fig. \ref{fig1}(c), a $B-T$ phase diagram is constructed in Fig. \ref{fig1}(d). The AFM transition is also reflected in a $\lambda$-shaped peak at $T_\mathrm{N}$ in the specific heat ($C_\mathrm{p}$) data presented in Fig. \ref{fig1}(e), where the phonon background is approximated by fitting the data above $T_\mathrm{N}$ using the Debye-Einstein model [black solid line in Fig. \ref{fig1}(e)]:
\begin{equation}
    C_\mathrm{ph}=\gamma T+9Na R(\frac{T}{\theta_\mathrm{D}})^3\int_{0}^{\theta_\mathrm{D}/T}\frac{x^{4}e^{x}}{(e^{x}-1)^{2}}dx+3N(1-a)R(\frac{\theta_\mathrm{E}}{T})^{2}\frac{e^{\theta_\mathrm{E}/T}}{(e^{\theta_\mathrm{E}/T}-1))^{2}},
\label{eq:Cp_Fitting}
\end{equation}
where $x = \hslash \omega / k_\mathrm{B} T$, $\omega$ is the phonon frequency, $\gamma$ is the Sommerfeld coefficient, $\theta_\mathrm{D}$ is the Debye temperature, $\theta_\mathrm{E}$ is the Einstein temperature, $a$ is a constant, $R$ is the ideal gas constant, and $N$ represents the number of atoms per unit cell ($N = 5$ for \ce{EuAl2Si2}). The extracted Debye temperature is $\theta_\mathrm{D} = 500$ K. The magnetic entropy is obtained by subtracting the phonon contribution from the total specific heat: $S_\mathrm{mag} = \int (C_\mathrm{p} - C_\mathrm{ph}) /TdT$. The saturated $S_\mathrm{mag}$ closely matches the theoretical value $R \ln(8)$ expected for $S = 7/2$ of Eu$^{2+}$, indicating a good estimation of the phonon background. The largest entropy release occurs below $T_\mathrm{N}$, with full saturation achieved only above 100 K, suggesting that short-range magnetic fluctuations extend up to approximately 100 K. These fluctuations significantly impact the electrical transport properties, as shown in Fig. \ref{fig1}(f). In zero magnetic field, the longitudinal resistivity ($\rho_{xx}$) decreases with cooling, exhibiting overall metallic behavior. However, below about 100 K, $\rho_{xx}$ increases slightly, likely due to enhanced electron scattering from increasing magnetic fluctuations. Below $T_\mathrm{N}$, magnetic fluctuations are quenched by the establishment of long-range AFM order, and metallic transport is recovered. Applying a magnetic field also suppresses magnetic fluctuations, resulting in metallic behavior down to temperatures below $T_\mathrm{N}$. At temperatures below about 30 K, $\rho_{xx}$ in 9 T increases with cooling, likely due to the localized cyclotron motion of carriers in strong magnetic fields.  

\begin{figure*}
\centering
\includegraphics[scale=0.215]{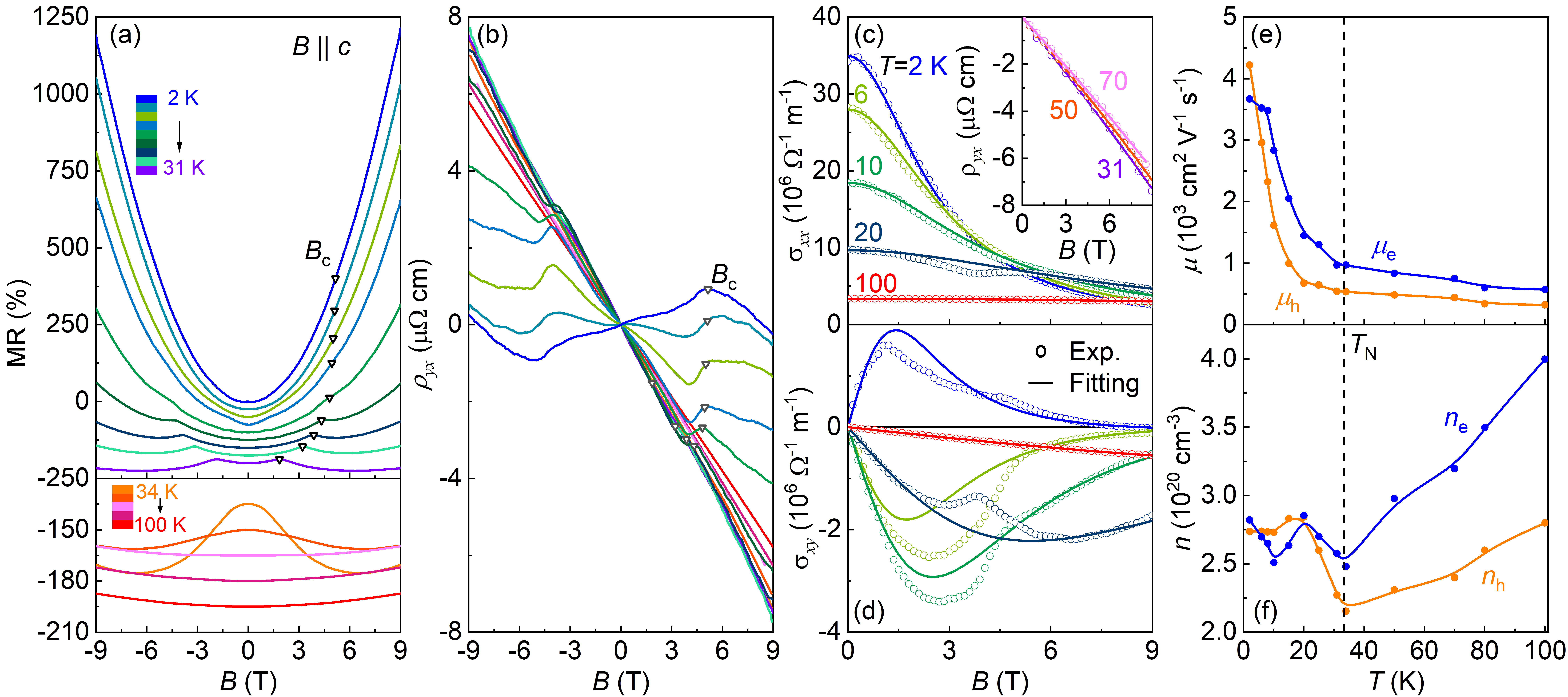}
\caption{  (a) Magnetoresistance (MR) and (b) Hall resistivity ($\rho_{yx}$) measured at selected temperatures. The data in (a) have been shifted vertically for clarity, and open triangles indicate the transition at $B_\mathrm{c}$. (c) and (d) Two-band fitting (solid lines) applied to the longitudinal ($\sigma_{xx}$) and transverse ($\sigma_{xy}$) magnetoconductivity data (open symbols). (e) and (f) Mobility ($\mu$) and carrier density ($n$) derived from the two-band fitting shown in (c) and (d). }
\label{fig2}
\end{figure*} 

In Fig. \ref{fig2}, we present the longitudinal magnetoresistivity (MR) and transverse Hall resistivity. At 2 K, a non-saturating MR $=[\rho_{xx}(B) - \rho_{xx}(0)] / \rho_{xx}(0)$ exceeding 1000\% is observed in 9 T. The magnetic transition at $B_\mathrm{c}$ is obscured by the rapid increase in MR and becomes noticeable only above 10 K when MR is moderate. In the intermediate temperature range between 30 and 70 K, negative MR is observed, attributed to field-induced suppression of magnetic fluctuations, consistent with the temperature-dependent resistivity shown in Fig. \ref{fig1}(f). The Hall resistivity ($\rho_{yx}$) exhibits non-monotonic behavior with respect to the magnetic field at low temperatures, and its sign changes around 4 K in small magnetic fields. These features are indicative of multiband transport involving competing electron- and hole-like carriers. In Figs. \ref{fig2}(c) and \ref{fig2}(d), we fit the longitudinal ($\sigma_{xx}$) and transverse ($\sigma_{xy}$) conductivities using two-band models (solid lines in Figs. \ref{fig2}(c) and \ref{fig2}(d)):
\begin{equation}
    \sigma_{xx}=\frac{\rho_{xx}}{\rho^2_{xx}+\rho^2_{yx}}=\frac{n_\mathrm{e}e\mu_e}{1+(\mu_\mathrm{e}B)^2}+\frac{n_\mathrm{h}e\mu_\mathrm{h}}{1+(\mu_\mathrm{h}B)^2},
    \label{eq3}
\end{equation}
\begin{equation}
    \sigma_{xy}=\frac{\rho_{yx}}{\rho^2_{xx}+\rho^2_{yx}}=\frac{-n_\mathrm{e}e\mu_\mathrm{e}^2B}{1+(\mu_\mathrm{e}B)^2}+\frac{n_\mathrm{h}e\mu_\mathrm{h}^2B}{1+(\mu_\mathrm{h}B)^2}.
    \label{eq4}
\end{equation}
Here, $n_\mathrm{e(h)}$ and $\mu_\mathrm{e(h)}$ denote the density and mobility of electron (hole)-like carriers. The simple two-band model effectively describes both $\sigma_{xx}$ and $\sigma_{xy}$ except near $B_\mathrm{c}$ in the AFM state. This deviation or extra contribution ($\sigma_{xy}^\mathrm{E}$) is attributed to stripe domain wall scattering induced domain wall Hall effect \cite{xia2023}, or skew scattering induced anomalous Hall effect (AHE) \cite{Chen_EuAlSi}. Giant extra Hall conductivity exceeding $1\times10^6$ $\Omega^{-1}\mathrm{m}^{-1}$ has been found in previous studies \cite{xia2023,Chen_EuAlSi}. By subtracting the fitted two-band results from the experimental data, a reduced $\sigma^\mathrm{E}_{xy}\sim2\times10^5$ $\Omega^{-1}\mathrm{m}^{-1}$ is obtained at 6 K and 4 T, like due to varying skew scattering strength or DW density in different samples \cite{Chen_EuAlSi}. The AHE contributed by magnetization is not considered in this analysis, as significant AHE is not expected from a collinear AFM state. Indeed, the magnetization induced AHE in \ce{EuAl2Si2} is negligible compared to the domain wall Hall and ordinary Hall contributions \cite{xia2023}. In the intermediate temperature range of 30 to 70 K, Eq. \ref{eq3} cannot account for the negative MR due to electron scattering from magnetic fluctuations. Thus, the two-band model was applied only to the Hall resistivity (see the inset in Fig. \ref{fig2}(c)). The extracted densities and mobilities are shown in Figs. \ref{fig2}(e) and \ref{fig2}(f). At temperatures above  $T_\mathrm{N}$, electron-like carriers dominate, as evidenced by the negative Hall coefficient (Fig. \ref{fig2}(b)). At lower temperatures, electron-hole compensation becomes apparent. The mobility of both carrier types increases rapidly below 20 K, with high mobilities of $\sim4000$ cm$^2$/Vs for both types. Our previous first-principles calculations, quantum oscillations (QOs), and angle-resolved photoemission spectroscopy (ARPES) measurements have revealed several electron and hole pockets near the Fermi energy \cite{xia2023}. In particular, a pair of Weyl points (WPs), located approximately 60 meV below $E_\mathrm{F}$, are formed by the linear crossing of hole bands near the $\Gamma$ point. A hole pocket, detected by QOs, is associated with the WP-derived bands \cite{xia2023}, likely contributing to the high mobility observed.

The thermoelectric properties of \ce{EuAl2Si2} are presented in Fig. \ref{fig:3}. The high-temperature negative Seebeck ($S_{xx}$) is consistent with the dominant electron-like transport observed in the Hall resistivity. In zero magnetic field, a slight kink in $S_{xx}$ near $T_\mathrm{N}$, indicates the AFM transition. Importantly,  $S_{xx}$ changes sign at approximately 10 K in 0 T, suggesting electron-hole compensation, similar to the behavior observed in Hall resistivity. Application of a magnetic field alters $S_{xx}$ only slightly, and $S_{xx}$ approaches zero around 10 K in 13 T. In contrast, a large Nernst signal exceeding 130 $\upmu$V/K is observed at low temperatures in 13 T [see Fig. \ref{fig:3}(b)]. In the high-field region, the Nernst effect peaks at a higher temperature  ($T_\mathrm{MH}\sim$ 8 K, 13 T), while in the low-field region, $S_{yx}$ reaches its maximum at a lower temperature ($T_\mathrm{ML}\sim$ 4 K, 3 T). Above these peak temperatures, $S_{yx}$ decreases rapidly with increasing temperature in both low- and high-field regions.  Notably, $S_{yx}$ decays as $1/T$ in 3 T [see Fig. \ref{fig:3}(b) inset]. These behaviors closely resemble the Nernst properties of the Weyl semimetal NbP \cite{Watzman_NbP}.  S. J. Watzman \textit{et al.} demonstrated that linearly dispersive bands can lead to a characteristic peak in $S_{yx}$, which occurs at $T_\mathrm{ML}\approx\sqrt{3}\epsilon_0/\pi k_\mathrm{B}$ and  $T_\mathrm{MH}\approx \epsilon_0/k_\mathrm{B}$ in the low-field and high-field regions, respectively ($\epsilon_0$ is the chemical potential at 0 K) \cite{Watzman_NbP}.  Above $T_\mathrm{ML}$, a $1/T$ power law temperature dependence of $S_{yx}$ is expected in the low-field case \cite{Watzman_NbP}. For \ce{EuAl2Si2}, the ratio $T_\mathrm{ML}/T_\mathrm{MH}=0.5$ is close to $\sqrt{3}/\pi$. These features suggest that the linearly dispersive bands forming the Weyl points play important roles in the Nernst thermopower of \ce{EuAl2Si2} at low temperatures. Note that in NbP, the Nernst signal is also 100 times that of the Seebeck signal, further indicating that similar physics is at play for both NbP and \ce{EuAl2Si2}.

\begin{figure*}
\centering
\includegraphics[scale=0.22]{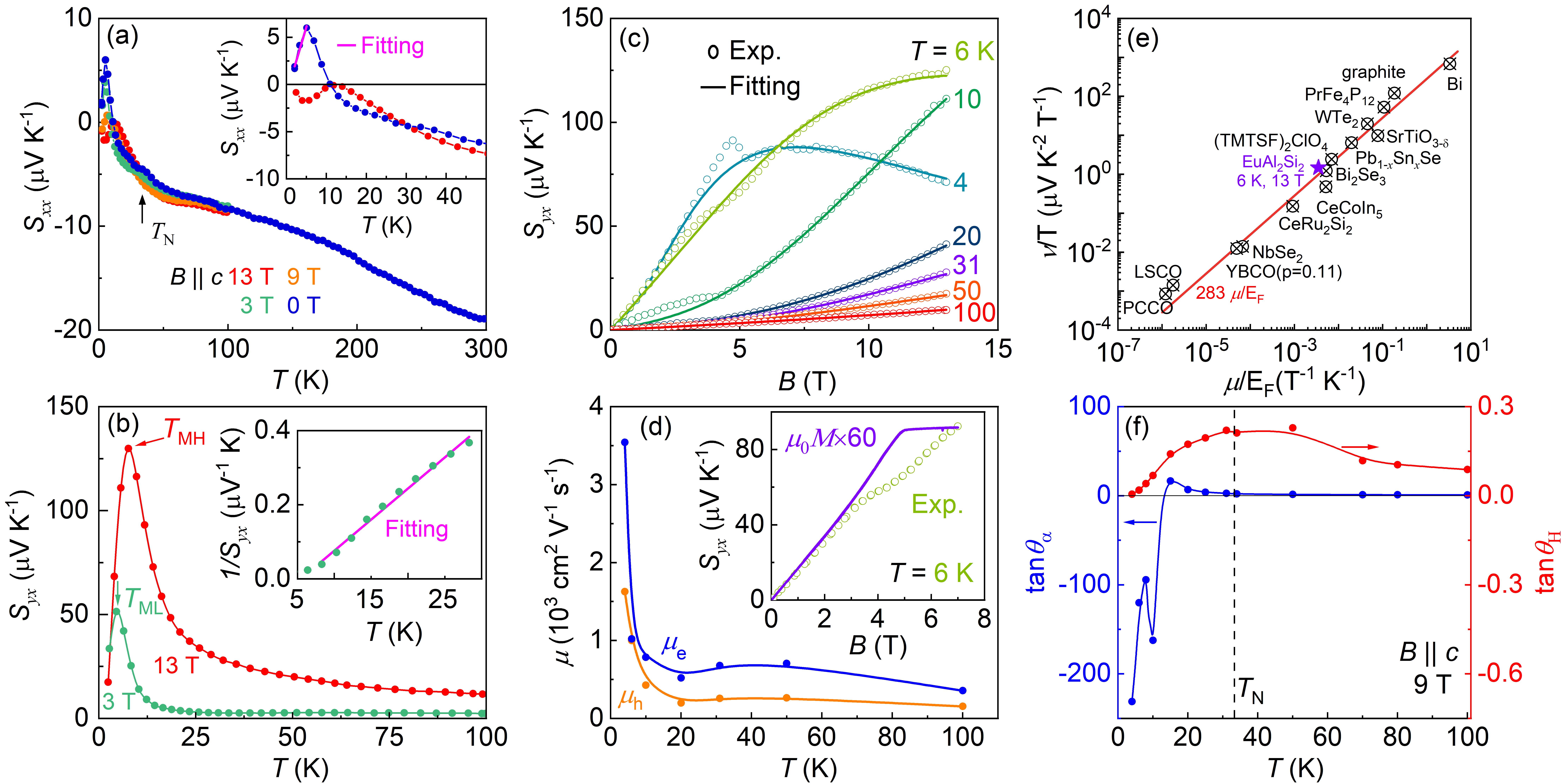}
\caption{(a) and (b) Temperature-dependent Seebeck ($S_{xx}$) and Nernst ($S_{yx}$) signals of \ce{EuAl2Si2}. The inset in (a) provides a zoomed-in view at low temperatures. The linear fitting (pink solid line) of $S_{xx}$ at low temperatures is used to extract the Fermi energy. (c) Two-band fitting (solid lines) applied to the Nernst data (open symbols). (d) Mobility of electron- and hole-like carriers obtained from the two-band fitting in (c). The inset in (d) compares the Nernst data with magnetization measured at 6 K, with the magnetization multiplied by a factor of 60 for comparison. (e) Nernst coefficient divided by temperature ($\nu/T$) plotted against the ratio of mobility to Fermi energy ($\mu/E_\mathrm{F}$) for various materials \cite{behnia2009,behnia2016}. The Nernst coefficient of \ce{EuAl2Si2} closely follows the semiclassical prediction (solid red line). (f) Temperature dependence of the Hall-like angles $\theta_\mathrm{H}$ and $\theta_\alpha$. }
\label{fig:3}
\end{figure*}

The isothermal  Nernst effect exhibits non-monotonic magnetic field dependence, indicative of multiband transport, as shown in Figs. \ref{fig:3}(c). The Nernst signal can be approximated using a semiclassical two-band model \cite{Liang_CdAs,Liang2013,Chen_CsVSb}:
\begin{equation}
   S_{yx}=N_\mathrm{e}\frac{\mu_\mathrm{e}B}{1+(\mu_\mathrm{e}B)^2}+N_\mathrm{h}\frac{\mu_\mathrm{h}B}{1+(\mu_\mathrm{h}B)^2}, 
\end{equation}
where $N_\mathrm{e}$ and $N_\mathrm{h}$ represent the Nernst contributions from electron-like and hole-like bands, respectively. The magnetization-induced anomalous Nernst contribution, $|S_{yx}^{A}|=|Q_S| \upmu_{0}M$ with $|Q_S|=0.05 - 1$ $\upmu$V/KT for typical magnets, is negligible [see Fig. \ref{fig:3}(d) inset], similar to the magnetization-induced AHE. The obtained mobility is of the same order of magnitude as that found in magnetoconductivity [see Fig. \ref{fig2}(e) and Fig. \ref{fig:3}(d)].

In the zero-temperature limit, the diffusive Seebeck contribution in the isotropic scattering approximation scales linearly with temperature $S_{xx}=\pi^2k^2_\mathrm{B}T/6eE_\mathrm{F}$ for a two-dimensional Fermi surface \cite{abrikosov1988fundamentals}. By linearly fitting the low-temperature $S_{xx}$, the Fermi energy of \ce{EuAl2Si2} is found to be $E_\mathrm{F}=8.7(5)$ meV, which is close to the value obtained from quantum oscillations \cite{xia2023}. According to the relation $\nu/T\sim283\mu/E_\mathrm{F}$ $\upmu$V/K$^2$T a large Nernst coefficient is expected in \ce{EuAl2Si2} due to its high mobility and small Fermi energy. As shown in Fig. \ref{fig:3}(e), the Nernst coefficient measured at 6 K and 13 T closely matches the theoretical expectation, indicating that high mobility and a small Fermi energy play crucial roles in producing the large Nernst effect observed in \ce{EuAl2Si2}. Additionally, nearly compensated electron-hole transport is essential for facilitating a large Nernst signal. This is because the Nernst effect is expected to be vanishingly small in a simple one-band metal due to Sondheimer cancellation \cite{Sondheimer,Wangyayu2001}: 
\begin{equation}
S_{yx}=S_{xx}\left(\tan\theta_{\alpha}-\tan\theta_\mathrm{H}\right)=\left.\frac{\pi^{2}}{3}\frac{k_\mathrm{B}^{2}T}{e}\left(\frac{\partial\,\tan\theta_\mathrm{H}}{\partial E}\right)\right|_{E_{F}},
\label{eq7}
\end{equation}
where $\theta_\mathrm{H}=\sigma_{yx}/\sigma_{xx}$ and $\theta_{\alpha}=\alpha_{yx}/\alpha_{xx}$ represent the Hall-like angles of electrical conductivity and thermoelectric Peltier conductivity ($\alpha_{yx}$, $\alpha_{xx}$). In a typical single-band metal, the electrical conductivity is generally energy-independent [($\left.\partial\,\tan\theta_\mathrm{H}/\partial E) \right|_{E_\mathrm{F}}=0$], resulting in the cancellation of $\theta_\mathrm{H}$ and $\theta_\alpha$, and leading to a vanishing $S_{yx}$ \cite{Wangyayu2001,behnia2009}. However, in a two-band system, this Sondheimer cancellation can be overcome via \cite{behnia2009,Gan_CsVSb,Mi2022,Mi2023}:
\begin{equation}
\begin{aligned}
S_{yx}=S_{xx}\left(\frac{\alpha_{yx}^\mathrm{h}+\alpha_{yx}^\mathrm{e}}{\alpha_{xx}^\mathrm{h}+\alpha_{xx}^\mathrm{e}}-\frac{\sigma_{yx}^{h}+\sigma_{yx}^\mathrm{e}}{\sigma_{xx}^\mathrm{h}+\sigma_{xx}^\mathrm{e}}\right),
\label{eq8}
\end{aligned}
\end{equation}
where the superscript $\mathrm{e}$($\mathrm{h}$) denotes the conductivity of electron (hole)-like bands. The impact of multiband effects on $S_{yx}$ is most significant when ideal electron-hole compensation is present, i.e., $\sigma_{yx}^\mathrm{h}=-\sigma_{yx}^\mathrm{e}$. In this case, $S_{yx}$  is maximized because the second term in Eq. \ref{eq8} becomes zero, while the first term remains nonzero since $\alpha_{yx}^\mathrm{h}$ and $\alpha_{yx}^\mathrm{e}$ share the same sign. As shown in Fig. \ref{fig:3}(f), both Hall-like angles are very small at high temperatures, while  $\theta_\alpha$ increases rapidly below 20 K. The cancellation between 
 $\theta_\alpha$ and $\theta_\mathrm{H}$ is avoided due to the presence of compensated bands, resulting in the large Nernst signal observed at low temperatures.   

In summary, a large Nernst effect is observed at low temperatures in \ce{EuAl2Si2}, with $S_{yx}$ reaching $130$ $\upmu$V/K at 8 K and 13 T. The substantial Nernst effect in \ce{EuAl2Si2} is driven by the high mobility of the linearly dispersing Weyl-point-derived bands, coupled with electron-hole compensation. This significant Nernst thermopower indicates that \ce{EuAl2Si2} could serve as a promising system for exploring Ettingshausen coolers at cryogenic temperatures.

\FloatBarrier
 We thank Ziji Xiang and Zengwei Zhu for fruitful discussions. We thank Guiwen Wang and Yan Liu at the Analytical and Testing Center of Chongqing University for technical support. This work has been supported by Chinesisch-Deutsche Mobilit\"atsprogamm of Chinesisch-Deutsche Zentrum f\"ur Wissenschaftsf\"orderung (Grant No. M-0496), the Open Fund of the China Spallation Neutron Source Songshan Lake Science City (Grant No. DG2431351H), Fundamental Research Funds for the Central Universities, China (Grant No. 2024CDJXY022). The work at ShanghaiTech University has been supported by the open project from Beijing National Laboratory for Condensed Matter Physics (2023BNLCMPKF002), the Double First-Class Initiative Fund of ShanghaiTech University, and Shanghai Sailing Program (23YF1426900).
\section*{AUTHOR DECLARATIONS}
\subsection*{Conflict of Interest}
 The authors have no conflicts to disclose.
\section*{DATA AVAILABILITY}
The data that support the findings of this study are available from the corresponding author upon reasonable request

\bibliographystyle{aapmrev4-1}

\end{document}